\documentclass[conference]{IEEEtran}
\IEEEoverridecommandlockouts
\usepackage{cite}
\usepackage{amsmath,amssymb,amsfonts}
\usepackage{algorithmic}
\usepackage{graphicx}
\usepackage{textcomp}
\usepackage{xcolor}
\usepackage{hyperref}
\usepackage{xurl}
\def\BibTeX{{\rm B\kern-.05em{\sc i\kern-.025em b}\kern-.08em
    T\kern-.1667em\lower.7ex\hbox{E}\kern-.125emX}}
\usepackage{tcolorbox}
\tcbuselibrary{breakable}
\usepackage{url}
\usepackage{mdframed}
\newtcolorbox{guidelinebox}[1]{
  colback=gray!5,
  colframe=black,
  fonttitle=\bfseries,
  title=Guideline~#1,
  breakable
}

\newtcolorbox{pitfallbox}[1]{
  colback=gray!5,
  colframe=black,
  fonttitle=\bfseries,
  title=Pitfall~#1,
  breakable
}

\newmdenv[
  linewidth=0.8pt,
  linecolor=black,
  roundcorner=2pt,
  innerleftmargin=6pt,
  innerrightmargin=6pt,
  innertopmargin=4pt,
  innerbottommargin=4pt,
  skipabove=0.8em,
  skipbelow=0.4em
]{simplebox}

\usepackage{xcolor}

\usepackage{xspace}

\newcommand{\lang}[1]{\textsc{#1}\xspace}

\newcommand{\Java}{\lang{Java}}
\newcommand{\JavaScript}{\lang{JavaScript}}
\newcommand{\TypeScript}{\lang{TypeScript}}
\newcommand{\Python}{\lang{Python}}
\newcommand{\CSharp}{\lang{C\#}}

\newcommand{\ready}[1]{\textcolor{black}{#1}}

\begin{document}

\title{GitHub Template Repositories: Served Domains, Maintenance, and Practitioner Guidelines}

\author{
\IEEEauthorblockN{
Leuson Da Silva\IEEEauthorrefmark{1},
Altaf Allah Abbassi\IEEEauthorrefmark{1},
Imen Trabelsi\IEEEauthorrefmark{1},
Paulo Borba\IEEEauthorrefmark{2},
Foutse Khomh\IEEEauthorrefmark{1}
}
\IEEEauthorblockA{
\IEEEauthorrefmark{1}Polytechnique Montr\'{e}al, Canada\\
Emails: \{leuson-mario-pedro.da-silva, altaf-allah.abbassi, imen.trabelsi, foutse.khomh\}@polymtl.ca
}
\IEEEauthorblockA{
\IEEEauthorrefmark{2}Centro de Informática, Universidade Federal de Pernambuco, Brazil\\
Email: phmb@cin.ufpe.br
}
}

\maketitle

\begin{abstract}
 Over time, GitHub has introduced different strategies for sharing reusable code artifacts. 
 In addition to fork-based reuse, \textit{template repositories} provide a distinct feature for generating new projects from scaffolding. 
 Although this feature has been available since 2019, little is known about the domains it supports, its maintenance characteristics, or the practices that guide practitioners for effective template design.
To address this gap, we conduct a large-scale empirical study of GitHub template repositories across the five most used programming languages. 
First, we mine and categorize templates to analyze the domains they serve, exploring the LLM-as-a-judge strategy. 
Next, we explore the reliability of templates by evaluating the associations between repository characteristics and activity, and quality-related issues (e.g., code smells, vulnerabilities, and security hotspots) through statistical analysis. 
Finally, we qualitatively analyze a representative subset of \textit{templates} to derive practical guidelines and recurring pitfalls for template design and management.
Our results show that \textit{Web Development} is the predominant domain across ecosystems, while maintenance and quality issues vary by programming language. 
We further find that high-quality templates tend to adopt established software engineering practices, while providing comprehensive documentation and clear guidance for use.
Overall, our findings offer empirical insights and actionable guidance to support practitioners in designing and adopting high-quality template repositories.
\end{abstract}


\section{Introduction}

Code and related artifacts are largely shared among developers and practitioners during collaborative software development mainly through version control systems, like \textit{Git} and \textit{Mercurial}.
Over time, these tools gained support from online services like GitHub and BitBucket, expanding the resources and features available to practitioners.
With advancements in GitHub and related features, they have shaped the current ways practitioners share code. 
For example, \textit{forks} allow developers to mainly extend repositories created and maintained by others, keeping parallel lines of development \cite{ren2018forks}, while \textit{pull requests} enable contributors to propose changes back to the original project, mediating collaboration and integration of external contributions \cite{gousios2014exploratory}.
On the other hand, \textit{template repositories}\footnote{Hereafter referred to as templates.} represent an alternative to share reusable code artifacts as starting points for new projects \cite{jackson2019repoTemplates}. 
While the former focuses on building software through collaborative development, with external contributors fixing and implementing new features \cite{jiang2017and}, the latter aims to efficiently speed up the initial steps of a project that shares relevant characteristics with the intended system, such as similar technologies or configurations, with less manual upfront configuration \cite{smith2024software, tu2022github}. 

Knowing that these features have been used for a while, we observe a discrepancy in what is known about them. 
While forks and pull request practices have received significant attention in prior research \cite{jiang2017and, zhou2020has}, \ready{templates have mostly been used to support research and educational activities \cite{google_scholar_search}, rather than to study how practitioners use them.} 
Furthermore, considering the expressivity of templates, with some of them becoming more popular than regular repositories, like a Microsoft repository focusing on Web Development for beginners \cite{microsoft_web_dev}, it is important to further understand the extent to which developers adopt templates, the purposes they serve, and the practices they are based on. 
Understanding these dimensions is essential to reasoning about where templates are most prevalent, how they differ across ecosystems, and which risks and benefits they may propagate to downstream projects. 
Without such understanding, both practitioners and researchers lack guidance on how to design, evaluate, and evolve templates in ways that effectively support software development practices.

Aiming to address this gap, we investigate (i) the domains that \textit{templates} serve for, (ii) how repository characteristics and activity relate to maintenance over time, and (iii) the practical guidelines and pitfalls that could support practitioners when creating and maintaining \textit{templates}. 
To this end, we conduct an empirical analysis of GitHub \textit{templates}, guided by three main research questions, across the five most widely used programming languages, according to the GitHub Octoverse ranking \cite{GitHubOctoverse2025}: \TypeScript, \Python, \JavaScript, \Java, and \CSharp.

First, for our RQ1, we mine \textit{templates} and, supported by LLMs, automatically categorize them to uncover their supported domains, drawing on taxonomies from prior work~\cite{sas2023gitranking,zanartu2022automatically}.
Our goal is to investigate and compare the domains in which practitioners employ them and examine how template usage differs across programming languages.
For RQ2, we focus on how repository characteristics and activity relate to maintenance and security issues, aiming to evaluate the reliability of templates as reusable artifacts and identify signals that may help practitioners anticipate potential downstream effort before adoption.
To this end, we collect a set of metrics capturing quality-related aspects (e.g., code smells, vulnerabilities, and security hotspots) and analyze their associations with repository characteristics (e.g., stars, forks, and commits) using a regression model.  
Finally, for RQ3, we select and analyze a set of repositories aiming to identify common patterns and derive actionable guidelines and pitfalls for practitioners when creating and maintaining their \textit{templates}.

Our results show that \textit{Web Development} is the main domain supported by \textit{templates} across all programming languages, whereas more specialized domains tend to be associated with specific ecosystems, such as \textit{Machine Learning/AI}, primarily supported by \Python.
For RQ2, we observe substantial variability in maintenance across \textit{templates} and programming languages.
While some \textit{templates} consistently exhibit low occurrence of bugs, code smells, and vulnerabilities, others concentrate multiple issues.
These differences vary across ecosystems, with no single pattern holding uniformly across programming languages.
Overall, the results show that maintenance and quality issues are ecosystem-dependent and cannot be treated as homogeneous signals across different programming languages.

For RQ3, as guidelines, we emphasize that \textit{templates} should adopt standard and good practices commonly found in regular GitHub repositories, such as automated pipelines and comprehensive documentation.
Beyond these practices, we also reinforce the importance of educating practitioners on properly using \textit{templates}, promoting reuse through families of templates, and aligning releases with supported technology versions to avoid duplication and improve sustainability.
As pitfalls, we highlight that repurposing regular repositories as \textit{templates} without proper adaptation, unclear or inactive maintenance status, and poor separation of concerns between templates and technologies can mislead practitioners and hinder effective template reuse.

Overall, our study makes the following contributions:
\begin{itemize}
    \item A set of guidelines and pitfalls that practitioners should consider when providing and selecting \textit{templates};
    \item A holistic view about the adoption of \textit{templates}, covering domains and maintainability; 
    \item A dataset of \textit{templates} covering the top 5 programming languages in GitHub.
    \item We make our data and results publicly available in our replication package \cite{online_appendix}.
\end{itemize}

\section{Data Collection}
\label{sec:data_collection}
\begin{figure}
    \centering
    \includegraphics[width=1\linewidth]{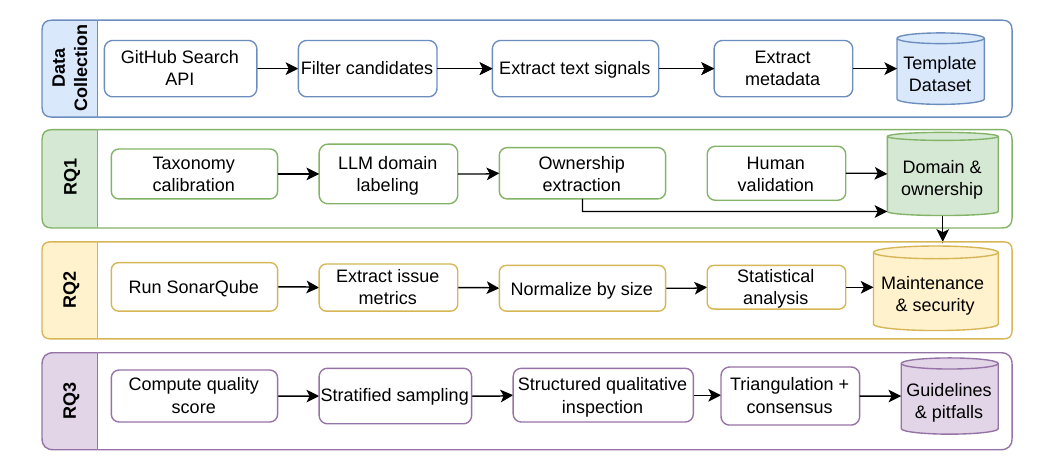}
    \caption{Overview of the research approach.}
    \label{fig:approach}
\end{figure}

In this section, we explain in detail our process to establish the sample of repositories to be analyzed, mainly covering the mining and filtering steps of GitHub \textit{templates}.
Similar to previous studies, we rely on the GitHub API \cite{cosentino2016findings} \ready{ and follow common GitHub mining guidelines to avoid duplicated, inactive, or non-representative repositories \cite{kalliamvakou2016depth,munaiah2017curating,borges2016understanding,zampetti2022empirical}.}
Figure~\ref{fig:approach} summarizes the overall workflow.

\subsection{Template Search Strategy}
\label{sec:search_process}
We rely on the GitHub API to collect our set of \textit{templates}. 
To the best of our knowledge, the API does not provide official documentation or explicit support for searching exclusively for \textit{templates} \cite{github_templates_search_unsupported}. 
During an in-depth inspection of the API behavior, however, we discovered that including the parameter \texttt{template:true} in the search query restricts the results to repositories explicitly marked as \textit{templates}.
To assess the reliability of this undocumented mechanism, we contrasted its results with those obtained through a keyword-based heuristic commonly adopted in prior studies. 
Specifically, we performed an exploratory search using terms frequently associated with \textit{templates}, including \textit{template}, \textit{boilerplate}, \textit{starter}, \textit{skeleton}, and \textit{scaffold}, appearing in repository names, descriptions, and README files. 
Our comparison showed that all repositories retrieved by the keyword-based approach were also captured by the \texttt{template:true} query, which additionally returned a substantially larger set of repositories.\footnote{We executed the keyword-based heuristic for comparison, but all its results were included, and substantially surpassed, by those retrieved with the \texttt{template:true} query.}
Based on this validation, we adopted the API-based approach relying on the \texttt{template:true} parameter for our data collection. 

\subsection{Filtering Criteria}
\label{sec:filtering_criteria}
We create a \Python script with a set of configuration parameters to ensure stable and reproducible queries (e.g., number of repositories per page, maximum pages per query, timeouts, and delay intervals). 
Regarding the adopted search parameters, we followed well-established guidelines for empirical studies on GitHub, ensuring that our queries targeted active, non-forked, non-archived, and public repositories \cite{munaiah2017curating,borges2016understanding,zampetti2022empirical}.
By excluding forked repositories, we aim to avoid duplication and promote a more diverse set of independent projects in our sample.
To capture active projects, we further exclude archived repositories and those that have not been updated in the past two years.
We only require repositories to have at least two stars as a minimal filter to remove trivial or toy projects, while still preserving a broad and diverse sample of \textit{templates}.
We did not impose any restriction on the number of forks, since \textit{templates} are intended to be reused directly rather than forked.
Finally, we did not apply any restrictions related to project structure (e.g., minimum number of files) or development history (e.g., minimum period of activity), as these aspects could limit our findings. 
Such criteria might be suitable for general-purpose repositories that evolve continuously, but they could bias the sample against valid \textit{templates}, which could be seen as incomplete when compared with regular repositories. 

Knowing that GitHub limits search results to the first 1{,}000 matches per query, we instruct our script to automatically divide the search space into adaptive star-count ranges. 
To estimate the number of repositories in a candidate range, we issue a lightweight search API request for the same query and star interval (with \texttt{per\_page=1}) and use the \texttt{total\_count} field returned by the API as an estimate of the number of matches. 
We then recursively split the star range (e.g., \texttt{stars:2..3}, \texttt{stars:4..7}, etc.) until each sub-range yields at most 1{,}000 results.
For each star range, our script queries the GitHub Search API using the constructed search parameters.
As the API returns at most 100 repositories per page, the script iterates through up to 10 pages per range, accumulating up to 1,000 repositories that meet the criteria.
We repeated this process for all programming languages evaluated in this study.
Finally, Table \ref{tab:mined_repos} presents the final distribution of mined templates (2nd column).

\begin{table}[t]
\centering
\caption{Number of mined repositories per programming language across RQ1–RQ3}
\begin{tabular}{l|r|r|r}
\hline
\multicolumn{1}{c|}{\textbf{\begin{tabular}[c]{@{}c@{}}Programming\\ Language\end{tabular}}} & \multicolumn{1}{c|}{\textbf{\begin{tabular}[c]{@{}c@{}}Mined \\ Templates\end{tabular}}} & \multicolumn{1}{c|}{\textbf{RQ1}} & \multicolumn{1}{l}{\textbf{\begin{tabular}[c]{@{}l@{}}RQ2/\\ RQ3\end{tabular}}} \\ \hline
\TypeScript                                                                                   & 6202                                                                                    & 5804                             & 6180                                                                           \\ \hline
\Python                                                                                       & 3363                                                                                    & 2628                             & 3356                                                                           \\ \hline
\JavaScript                                                                                   & 3708                                                                                    & 3176                             & 3368                                                                           \\ \hline
\Java                                                                                         & 938                                                                                      & 740                               & 262                                                                             \\ \hline
\CSharp                                                                                          & 569                                                                                      & 470                               & 215                                                                             \\ \hline 
\textbf{Total} & \textbf{14780} & \textbf{12818} & \textbf{13381} \\ \hline
\end{tabular}
\label{tab:mined_repos}
\end{table}

\section{RQ1: What domains and ownership patterns characterize templates?}
\label{sec:rq1}

\subsection{Motivation}
\label{sec:rq1_motivation}
When \textit{templates} are created, they are typically designed to address specific goals using particular technologies. 
These goals often reflect the needs of particular domains, considering the diversity of repositories and ecosystems on GitHub. 
Accordingly, identifying the domains served by \textit{templates} in our final classification of the analyzed sample allows us to understand where \textit{templates} are most prevalent and where gaps exist.
Furthermore, this knowledge also reveals how domain-specific requirements (frameworks, tools, or patterns) might influence template design and maintenance.
In addition to the domain, ownership, whether a template is created by an individual developer or an organization, also shapes its purpose, quality, and expected use. 
Understanding these ownership patterns provides complementary insights into who drives template creation and how \textit{templates} propagate across the ecosystem.

\subsection{Approach}
\label{sec:rq1_approach}
With the final list of projects selected (Table \ref{tab:mined_repos}), we proceed to infer the domains these projects serve.
\ready{Since we are not aware of prior studies using LLMs to classify the domains of GitHub repositories, we use prior LLM-based SE studies \cite{abbassi2025unveiling,sollenberger2024llm4vv} as motivation for employing LLMs as support tools in our classification and judgment tasks.}
\ready{To design and validate the domain taxonomy used in this process, we first conducted a manual analysis on a calibration subsample. Specifically, we randomly selected 100 projects from the \Python sample, and one researcher manually categorized them according to domain classifications proposed in prior work~\cite{sas2023gitranking,zanartu2022automatically}. This initial analysis showed that most repositories fit into eight existing categories.}
However, several projects could not be adequately captured by the original taxonomy.
As a result, six additional domains emerged and were incorporated into the final classification schema: \textit{Automation/DevOps, Infrastructure/Cloud, Research, Data Management, Robotics/IoT, and Data Apps/Visualization}.

Once the categories were defined, concise definitions for each category were created to guide the LLM-based classification.
\ready{Guided by recent reporting guidelines for empirical SE studies involving LLMs~\cite{baltes2025guidelines}, we report the LLM role, model configuration, prompt inputs, output format, and human validation procedure; additional details are available in our online appendix~\cite{online_appendix}.}
\ready{For each repository, the prompt included the project description, topics, and README content, when available. 
The model was instructed to (i) summarize the project, (ii) assign exactly one category according to the predefined taxonomy, (iii) justify its decision by briefly explaining the rationale behind the classification, and (iv) report its confidence in the classification using a score from 1 to 5. 
If a project did not fit any of the predefined categories, the model was allowed to propose a new one. 
The output was requested in JSON format, including the fields \texttt{category}, \texttt{reasoning}, and \texttt{confidence}.}
Two models were selected based on their capabilities (\texttt{GPT-4o-mini} and \texttt{DeepSeek-V3.2}), employed with a temperature setting of 0.2, using our manually classified templates as a reference.
\texttt{DeepSeek-V3.2} achieved higher agreement with the manually labeled sample (91/100) compared to \texttt{GPT-4o-mini} (74/100). 
Therefore, we selected \texttt{DeepSeek-V3.2} for the remaining classification tasks.
During this process, we observed that some project information was not available in English.  
To ensure consistency in the analysis, we implemented preprocessing routines using \texttt{langdetect}\cite{langdetect_1_0_9} for language identification, which filtered out non-English projects based on the language detected in their README files, retaining only English repositories for classification (see Table~\ref{tab:mined_repos}, 3rd column).
\ready{During this process, the LLM proposed only three new domains among more than 12,000 classifications. 
After manual inspection, we found that these labels were technology-specific refinements of existing categories rather than missing high-level domains. 
We therefore mapped them to the closest predefined category in our taxonomy.}

As LLMs were used to classify template domains, this approach entails additional computational and financial costs.
In total, the LLM-based classification incurred a cost of approximately \$6~USD.
To evaluate the accuracy of the LLM’s classifications, two researchers independently categorized a different random subsample of projects, establishing a ground truth for comparison. 
Such a manual validation was performed on a statistically representative sample of 192 repositories (95\% confidence level with a 7\% margin of error). 
\ready{Two authors independently labeled the sampled repositories following the same protocol and using the resulting extended taxonomy.} 
In cases of disagreement, the authors discussed the repositories and reached consensus without consulting the labels produced by the LLM.
\ready{When comparing these consensus labels against the automated classifications, we observed that the LLM differed from the human consensus in 28 out of 192 cases, corresponding to 85.4\% agreement, with Cohen’s $\kappa=0.82$ specifically measuring the agreement between the automated classifications and the final human-consensus labels.
}

Finally, to analyze ownership patterns, for each template, we extracted the owner type, distinguishing whether it was maintained by an individual user or an organization (all mined templates, see Table \ref{tab:mined_repos}, 2nd column). 
After collecting this information, we computed the frequency of each ownership type across all templates. 
Such an analysis allows us to quantify the distribution of templates between individual and organizational owners and to identify overarching trends in their prevalence.

\subsection{Results}
\label{sec:rq1_results}

\begin{figure}[t]
\includegraphics[width=\linewidth]{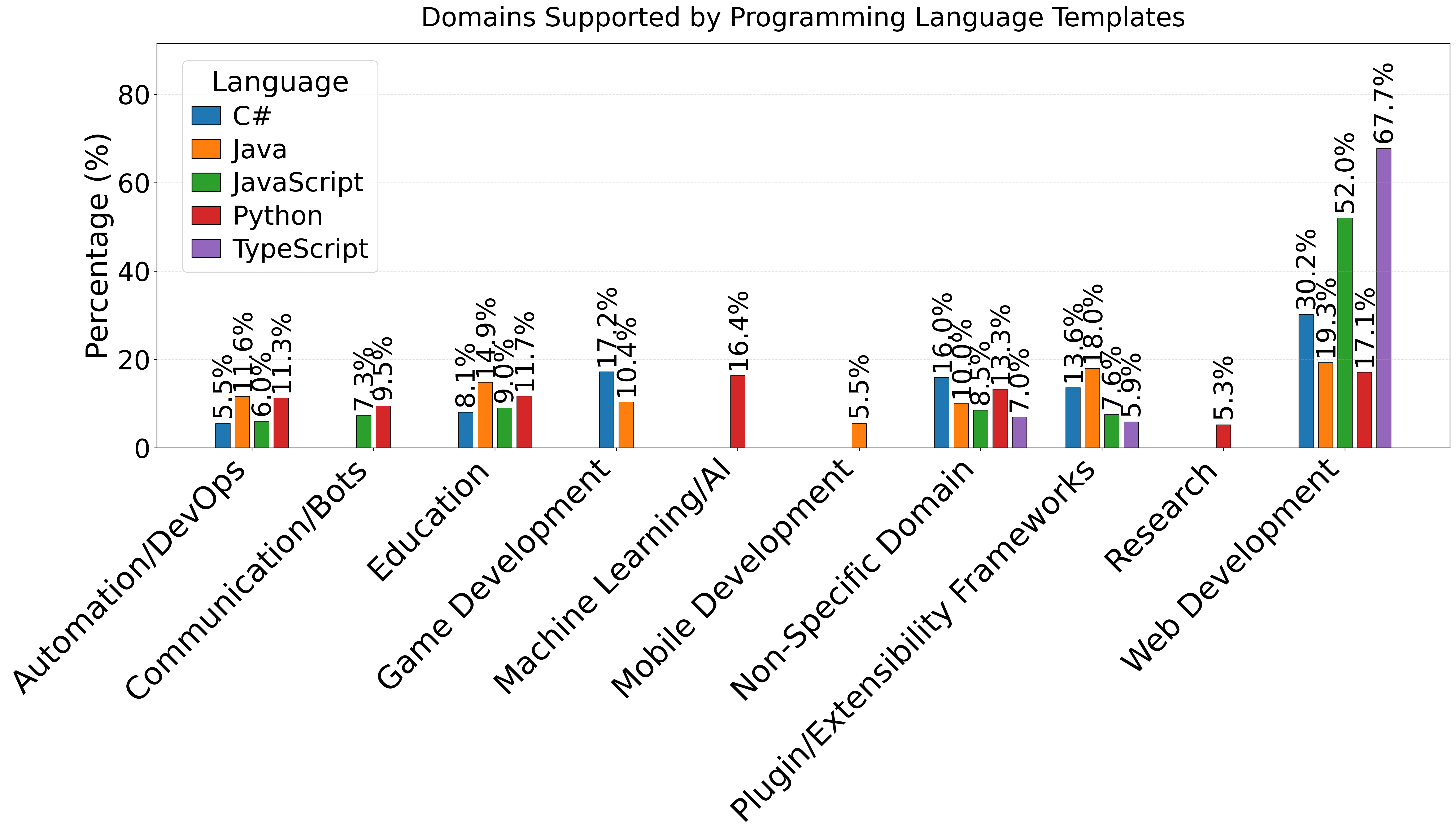}
  \caption{Distribution of Domains supporting GitHub Templates.}
  \label{fig:usage_domain}
\end{figure}

\subsubsection*{Domain Analysis}
Figure \ref{fig:usage_domain} presents the distribution of the domains associated with the templates across the five programming languages evaluated here.
We report only categories accounting for more than 5\% of the repositories. Full details are provided in the online appendix \cite{online_appendix}.
Overall, we can observe a clear dominance of \textit{Web Development} templates, which represent the most dominant domain across all programming languages. 
Those results, specifically for \TypeScript (67.7\%) and \JavaScript (52\%), are expected since these languages are heavily used in front-end and full-stack projects, highly supported by web frameworks (React, Next.js, Angular) \cite{di2025investigating, GitHubOctoverse2025}.
For the remaining domains, we observe a more balanced distribution, highlighting \textit{Non-specific Domain}, \textit{Plugin/Extensibility}, and \textit{Education} templates.
While the \textit{Non-specific Domain} category provides generic, domain-agnostic software templates for a wide range of software development purposes, the \textit{Plugin/Extensibility} category supports scaffolding projects designed to extend existing resources or frameworks.
Meanwhile, \textit{Education} templates focus on offering a structured foundation for projects whose primary goal is teaching or learning, such as course materials, assignments, or programming competitions.
These templates often include configurations and structures that facilitate integration with larger platforms (e.g., IDE extensions, web frameworks, CI/CD, and general tools), enabling practitioners to streamline the development of plugins, modules, or reusable components.

Different from \textit{Regular Projects}, \textit{Automation/DevOps} templates provide additional support for various types of automation, including testing environments, CI/CD configurations, and deployment workflows.
The remaining domains exhibit either isolated highlights or uniformly small proportions.
For example, we observe a high percentage of \textit{Machine Learning/AI} templates in \Python, while a similar trend appears for \textit{Game Development} in \CSharp and \Java.
Such an observation aligns with the widespread use of \Python for ML projects on GitHub\cite{raschka2020machine}, as well as the well-established support for game development in \CSharp and \Java, considering key game engines, like Unity and JMonkey \cite{sharif2021game,GitHubOctoverse2025}.

\subsubsection*{Programming Languages Influence} While some programming languages concentrate their presence in a restricted number of domains (\TypeScript and \JavaScript), we observe that \Python is the top language covering all domains with substantial representation.
Such an observation highlights the versatility and broad applicability of \Python, whose ecosystem supports diverse software development needs ranging from web and data management to automation and machine learning.
In contrast, the restricted presence of other languages across domains underscores their more specialized or framework-oriented nature, suggesting that certain ecosystems foster narrower patterns of reuse and template creation.
This finding calls attention to the need for further investigation into how language-specific communities shape the dissemination and generalizability of templates, and also the imminent need for templates in these domains.

\begin{figure}[t]
  \centering
  \includegraphics[width=\linewidth]{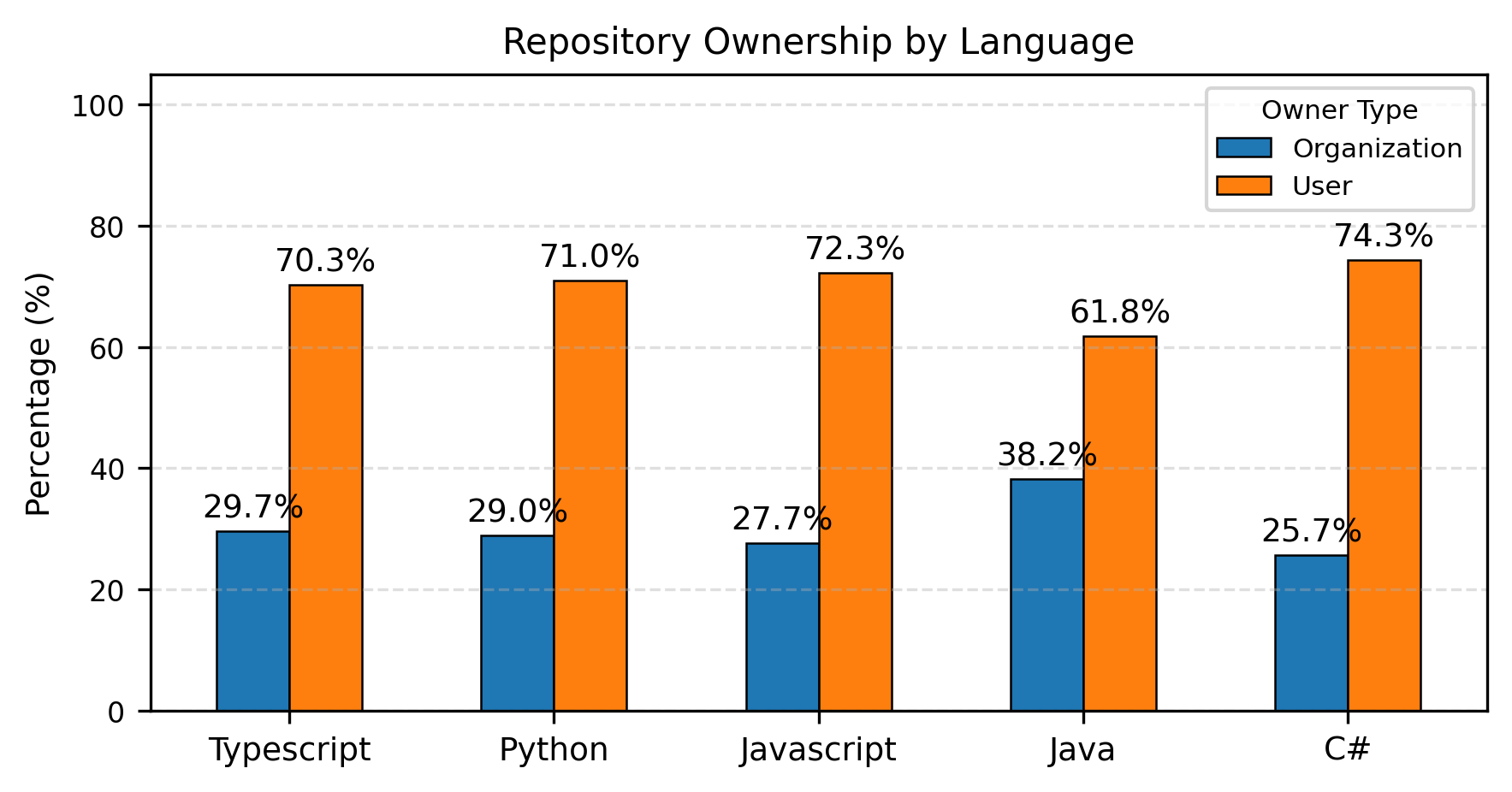}
  \caption{Distribution of repository ownership types (Organization vs User) across programming languages.}
  \label{fig:ownership_distribution}
\end{figure}

\subsubsection*{Template Ownership}
Figure~\ref{fig:ownership_distribution} presents the distribution of ownership on the evaluated templates.
On GitHub, a repository can be owned either by an organization (typically representing teams or companies) or by an individual user account \cite{zoller2020topology}.
Overall, we observe that individual users own the majority of templates, accounting for roughly two-thirds of the total. 
Although these ownership types primarily differ in purpose, we can infer common motivations for sharing templates.
Specifically, organizations tend to share templates to promote standardization and streamline internal development processes, whereas individual users often publish templates to support the open-source community, demonstrate best practices, or showcase reusable project structures.

\section{RQ2: How do repository characteristics and activity over time relate to maintenance?}
\label{sec:rq2_results}

\subsection{Motivation}

\textit{Templates} are primarily designed to be reused as starting points for new projects, reducing initial setup effort. 
Unlike traditional repositories that evolve independently \cite{rehman2022newcomer}, adopting \textit{templates} may directly propagate their structure, dependencies, and potential flaws to downstream projects \cite{wu2023understanding, lazarine2022exploring}. 
Consequently, the maintenance quality of templates is particularly important, as defects or outdated dependencies may be replicated across multiple projects and contribute to long-term technical debt \cite{he2023automating, cogo2021deprecation}. 
In practice, developers often rely on observable repository signals (e.g., activity, popularity, or ownership) when selecting templates, yet it remains unclear whether such signals relate to template reliability. 
Understanding the maintenance characteristics of templates is therefore essential to assess the risks they may pose across the ecosystem \cite{palomba2018beyond, kumar2024comprehensive}. 
In this RQ, we investigate how repository characteristics and activity relate to the presence of maintenance and security issues in templates, aiming to identify signals that may help practitioners assess template reliability before adoption.

\subsection{Approach}
When evaluating the occurrence of maintenance issues in GitHub repositories, previous studies have focused on several dimensions, including code smells, bugs, reliability issues, and security vulnerabilities. 
These studies typically rely on static analysis tools to detect, quantify, and report such issues at scale \cite{lefever2021lack,lenarduzzi2020some}.
Motivated by these studies, we aim to investigate the occurrence of quality issues specifically within the context of templates. 
Specifically, we treat the number of maintenance and security issues as dependent variables, and analyze their associations with repository characteristics and activity indicators.

Since our study covers five different programming languages and requires collecting comparable results across them, it was essential to adopt a tool with broad language support and consistent reporting capabilities. 
For this reason, we selected SonarQube \cite{sonarqube_community_2025}, a widely used and industry-standard static analysis platform. 
To operationalize the analysis, our scripts handle all the required steps to run SonarQube and collect metrics.
For each template, the script clones the latest available repository's version, invokes SonarQube on this snapshot, and extracts the resulting quality metrics: \textit{bugs}, \textit{code smells}, \textit{security hotspots}, \textit{general} and \textit{informational vulnerabilities}, and their associated \textit{severity levels}.
For exploratory analyses, we normalized issue values by size (per KLOC) to enable fair comparisons across repositories. 
For compiled programming languages, successfully building the repository is a prerequisite for running SonarQube.
\ready{However, a subset of \Java and \CSharp repositories could not be analyzed due to missing dependencies, configuration issues, or build failures. 
Specifically, 676 \Java repositories and 354 \CSharp repositories were excluded from the corresponding analyses.}
For the remaining programming languages, a small number of runs still failed due to SonarQube execution issues (e.g., timeouts or repository-specific configuration errors).
These repositories were excluded, resulting in minor reductions of the final samples, as reported in Table~\ref{tab:mined_repos} (4th column).

To examine the association between repository characteristics and quality issues, we considered a regression model, allowing us to control for multiple factors simultaneously and assess their independent effects.
Such a combination of methods provides a more rigorous understanding of which repository attributes are meaningfully associated with the observed maintenance issues.

First, we merged the metadata extracted from GitHub (RQ1) with the quality metrics obtained from SonarQube: number of \textit{recent commits} (last 12 months), \textit{stars}, and \textit{forks}. 
From these raw fields, we computed additional explanatory variables: \textit{repo\_age\_days} (repository's age in days), \textit{staleness\_days} (days since the last update), and logarithmic transformations of stars and forks to reduce skewness in popularity distributions. 
We also encoded the \textit{repository owner} as a binary variable, indicating whether the owner was an organization or an individual. 
For regression modeling, we used raw issue counts and incorporated repository size as a logarithmic offset. 
Finally, we consider the Negative Binomial regression model, which allows us to estimate the independent contribution of each predictor while controlling for multicollinearity. 
We also remove extreme outcome values to reduce the influence of outliers and assess the multicollinearity among predictors using variance inflation factors (VIF) before model fitting.
Finally, we also performed a Mann–Whitney U test to compare issue distributions between organization- and user-owned templates. 
\ready{Since our analyses involve multiple tests across programming languages, predictors, and outcomes, we applied the Benjamini--Hochberg false discovery rate correction within each programming language. 
} 
Together, these steps provide a comprehensive statistical characterization of how different repository attributes relate to the quality issues detected in templates. We adopt a significance level of $\alpha = 0.05$ for all statistical tests.

\subsection{Results}

Overall, we observe a strong dispersion, indicating that quality issues in templates are sparse, highly skewed, and dominated by a few repositories with disproportionately high issue counts.
When considering individual metrics, an average of 39–54\% of templates per language report zero issues, with the vulnerability metric exhibiting the highest sparsity (approximately 84–96\% of repositories reporting zero vulnerabilities). 
Skewness coefficients range on average from approximately 6 to over 46, indicating long-tailed distributions dominated by a few repositories. 
These characteristics violate normality assumptions, motivating our reliance on Negative Binomial regression models.

\ready{\textbf{Ownership differences.}
Organization-owned templates generally exhibit slightly lower issue densities, except for \TypeScript, but the effects are mostly negligible or small.
For \JavaScript and \Python, significant differences mostly indicate lower issue densities for organization-owned templates, with negligible effects ($r_{rb}\approx -0.02$ to $-0.10$), except for \Python critical vulnerabilities ($r_{rb}=-0.12$).
\Java and \CSharp show small negative effects for vulnerability density ($r_{rb}=-0.14$ and $r_{rb}=-0.11$), suggesting fewer vulnerabilities per KLOC in organization-owned repositories.
For \TypeScript, significant differences appear only for critical and informational vulnerability densities, both with negligible positive effects ($r_{rb}=0.05$ to $0.07$).
Overall, ownership type explains only a minor portion of variability in vulnerability density and maintenance indicators.
}

\textbf{Regression Analysis.}
We observe that \ready{repository-level metadata is associated with maintenance and security issues, but in language- and outcome-dependent ways.
The strongest evidence is observed for \JavaScript and \TypeScript, where \textit{log\_forks} is positively associated with bugs, code smells, general and informational vulnerabilities, security hotspots, and vulnerability severity levels, with IRRs from 1.20 to 2.29.
This suggests expected issue counts ranging from 20\% to 129\% higher for more diffuse templates in these ecosystems.
In contrast, for \Python, the only association involving \textit{log\_forks} is negative for bugs (IRR = 0.43), indicating 57\% lower expected bug counts.
Thus, repository diffusion relates to issue counts in an ecosystem-specific rather than universal way.
}

\ready{Recent development activity is associated with issue counts only in specific ecosystems, with varying direction and magnitude.
In \TypeScript, \textit{commits\_last\_12m} is positively associated with most evaluated issues, but with IRRs close to 1.00, indicating marginal increases of about 0.10\% to 0.35\% per additional commit.
In \CSharp, recent commits are positively associated only with security hotspots (IRR = 1.006), while \JavaScript shows mixed and very small corrected associations.
Thus, recent activity is a statistically detectable but modest signal, whose practical effect may only become meaningful cumulatively in highly active repositories.
}

\ready{
Finally, organization-owned templates are associated with fewer issues in some cases (\textit{owner\_is\_org}), including \Java vulnerabilities (IRR = 0.14), several \JavaScript outcomes (IRR = 0.31--0.73), and \CSharp bugs (IRR = 0.24).
Conversely, they are associated with more bugs in \Java and \TypeScript (IRR = 2.98 and 2.79), as well as more \TypeScript code smells and selected vulnerability severity levels (IRR = 1.25--1.62).
Thus, ownership likely reflects ecosystem-specific practices, scale, and reporting processes, and should not be interpreted as a uniform quality signal across languages.}

\section{RQ3: What actionable insights can support practitioners in creating, maintaining, and adopting templates effectively?}
\label{sec:rq3}

\subsection{Motivation}

\textit{Templates} are expected to accelerate project setup and promote reuse by providing structured scaffolding for new projects. However, as observed in RQ1 and RQ2, \textit{templates} span diverse ecosystems and may exhibit maintenance and security issues. This highlights the importance of understanding how \textit{templates} should be designed and maintained in practice. Therefore, beyond characterizing template repositories, it is important to derive actionable insights that can guide practitioners in the effective creation, management, and adoption of \textit{templates}. In RQ3, we aim to extract practical guidelines and identify recurring pitfalls that can help practitioners better design, maintain, and use \textit{templates}.

\subsection{Approach}
To identify actionable insights for practitioners, we performed a targeted qualitative analysis of a representative subset of templates. 
For that, first, we computed a composite quality score for each repository by combining the multiple maintenance and security metrics from RQ2, further normalizing them by project size (LOC, files, or functions).
Then, each normalized metric was transformed into a percentile rank to ensure robustness to skewed distributions and outliers, as previously observed in our sample. 
The composite score corresponds to the average percentile rank across all available metrics, with lower values indicating better overall quality.
Based on this score, we selected three complementary groups of repositories per programming language: best (20, lowest composite scores), worst (20, highest composite scores), and mixed repositories (10), characterized by high variability across quality dimensions. 
To mitigate sampling bias, selection was stratified by ownership type (organization vs. individual) and repository popularity (stars). 
This sampling strategy enables a focused yet representative qualitative examination of recurring quality patterns and practices that can support practitioners when selecting, reusing, and creating new templates.

\ready{To support a systematic qualitative analysis, we used predefined CSV templates to guide data extraction for each repository. The templates captured: (i) the repository goal and whether it was aligned with the target repository goal, (ii) quality and maintenance practices, and (iii) observed strengths, issues, and candidate actionable insights. 
Before the main analysis, two researchers conducted a calibration phase on a small random subset of templates outside the selected sample to refine the extraction fields, align the unit of analysis, and define the criteria for identifying actionable observations. 
Next, both researchers independently analyzed the sampled repositories and recorded their observations in the CSV files. 
The resulting observations were compared and grouped into recurring patterns. 
Patterns associated with desirable practices were consolidated as \textit{guidelines}, while recurring problematic practices were consolidated as \textit{pitfalls}. 
Disagreements were resolved by re-examining the repositories and RQ2 metrics. 
Finally, the agreed set of guidelines and pitfalls was reviewed by the remaining authors.}

\subsection{Results}
First, we discuss the set of guidelines observed as good practices, and next, the set of pitfalls.

\subsubsection{Guidelines} 

\begin{simplebox}
\textbf{G1: Integrate maintenance and quality practices}\end{simplebox}

Templates reported with the best scores in our analysis consistently adopted established software engineering practices. 
In particular, they provided ready-to-use project setups and integrated common quality mechanisms, such as linting, testing at appropriate levels, CI/CD pipelines, hooks, automation tools, and bots, for different purposes, like dependency updates and code review. 
These practices not only help promote maintainability and improve the quality of downstream projects instantiated from the templates, but may also serve an educational role by introducing good practices to new practitioners.

\begin{simplebox}
\textbf{G2: Provide comprehensive/accessible documentation}
\end{simplebox}
 
Beyond explaining the project’s purpose, templates must guide users in understanding, customizing, and extending the provided structure to fit their specific needs. 
To this end, effective documentation can take multiple forms, including step-by-step guides, code snippets with explanatory descriptions, links to external resources, and creation of wikis \cite{rife2}.
We also observe some templates providing documentation in different languages, supporting reuse while avoiding the need to create new templates to broadly support the same initial goal \cite{lgwk42}.

\begin{simplebox}
\textbf{G3: Educate users on how to GitHub Template Feature}
\end{simplebox}

Although GitHub’s templating feature is not new, we observe that several repositories provide inaccurate or incomplete instructions on how to properly use it. 
In particular, users are often instructed to clone the repository directly, rather than using the \textit{Use this template} button \cite{gpauloski}. 
Templates should explicitly guide users toward the correct instantiation mechanism and update their documentation accordingly, especially when existing repositories are later converted into templates.

\begin{simplebox}
\textbf{G4: Communicate evolution through a roadmap}
\end{simplebox}

GitHub repositories usually provide roadmaps to communicate upcoming changes and planned features.
Templates can equally benefit from this practice, as a roadmap helps users understand how the template is expected to evolve and how it can be extended over time.
This is particularly valuable for templates targeting the \textit{Education} domain, where roadmaps can guide learners by informing tasks to be performed and expected outcomes \cite{The-Marcy-Lab-School,hezean}.

\begin{simplebox}
\textbf{G5: Organize templates as a reusable family}
\end{simplebox}

Templates can be designed as a reusable family \cite{montalvillo2015tuning}, where new templates are derived from existing ones to preserve and propagate previously established good practices while introducing technology-specific variations.
We observe this approach when practitioners define a default template and subsequently use it as the basis for creating additional templates targeting different technologies, ecosystems, and programming languages \cite{thiagoroock_microsservice}.

\begin{simplebox}
\textbf{G6: Align technology versions with releases}
\end{simplebox}

Our analysis shows that many templates are created to support the use of specific technologies, often alongside libraries or frameworks provided by organizations or individual developers.
As these technologies evolve, templates should explicitly track the supported technology versions and associate them with corresponding template releases \cite{Jaxelr_VueSimpleTemplate}.
Providing clear release information, for example, through versioned releases, 
allows practitioners to identify compatible template versions while supporting reproducibility over time \cite{BunnyWay}.

\begin{simplebox}
\textbf{G7: Encourage community and social support}
\end{simplebox}

Collaborative software development inherently relies on community engagement through discussion, contribution, and shared problem-solving.
For templates, this engagement can be impacted by how practitioners consume, adapt, and extend the provided templates.
In our analysis, we observed some initiatives aimed at fostering active communities around templates, such as dedicated communication channels (e.g., Discord servers) to support users.
Similarly, some template providers include calls for sponsorship, helping sustain long-term maintenance in ways comparable to traditional GitHub repositories \cite{SpongePowered}.

\subsubsection{Pitfalls}

\begin{simplebox}
\textbf{P1: Turning non-templates into templates}
\end{simplebox}

Many repositories labeled as templates were originally designed as full implementations, such as end-to-end applications, libraries, or frameworks \cite{lschoe}.
While some templates may provide useful examples of how to use specific technologies, non-templates often lack the scaffolding, guidance, and extensibility expected from a template.
As a result, practitioners should avoid turning regular repositories into \textit{templates} unless they are explicitly adapted to support reuse, learning, and extension. 
This issue can be mitigated by following Guideline~G3, which helps ensure that repositories are intentionally designed and used as reusable starting points rather than being repurposed from full implementations.

\begin{simplebox}
\textbf{P2: Unclear or inactive repository status}
\end{simplebox}

Several inactive templates fail to clearly communicate their current status to practitioners \cite{learnthisacademy}.
To avoid misleading users, template providers should explicitly signal inactivity, for example, by archiving the repository or clearly documenting its status.
While we observed some cases where inactivity was mentioned in the README, such information is less visible and may be overlooked compared to GitHub’s native archival mechanisms \cite{EnderJs_CreatorGL}. This issue can be mitigated by following Guidelines~G4 and ~G6, which help communicate the evolution and maintenance status of templates over time.

\begin{simplebox}
\textbf{P3: Multiple templates in a single repository}
\end{simplebox}

GitHub repositories are typically designed as single entities with a clear and focused goal, and \textit{templates} are no exception.
However, we observed 
\textit{templates} targeting different programming languages or ecosystems within the same repository, distributed across folders, or even separate \textit{branches} \cite{CleanroomMC}.
Such an organization increases complexity and can negatively impact discoverability, maintenance, and reuse.
To mitigate this issue, practitioners should identify the common core across such variants and structure them as a family of templates, following the approach discussed in Guideline~G5.

\begin{simplebox}
\textbf{P4: Mixing technology and template concerns}
\end{simplebox}

Repositories hosting technologies on GitHub typically serve as primary points for implementation, documentation, issue tracking, and releases.
\textit{Templates}, however, address a different concern: providing reusable scaffolding rather than serving as the primary implementation artifact.
During our analysis, we observed cases where technology implementations and their corresponding templates were colocated within the same repository \cite{abersheeran}.
Such an organization can be confusing for practitioners and may hinder long-term evolution by mixing product-level changes with template-specific concerns.
Guideline~G5 helps address this issue by promoting the organization of templates as a family across technologies.

\begin{simplebox}
\textbf{P5: Empty templates and missing documentation}
\end{simplebox}

While \textit{templates} are expected to provide reusable scaffolding, we observe some empty repositories, where no files were available, only the structure of folders.
Similarly, some \textit{templates} also lack basic documentation, including a repository description or a README file.
These \textit{templates} provide insufficient guidance and offer limited support for practitioners aiming to understand, adapt, or extend them.
\textit{Template} providers should include at least minimal functional content and documentation, as well as an overview of the repository structure, tooling (e.g., pipelines), and, when appropriate, high-level graphical representations \cite{NikolayIT}. 
Guideline~G2 helps ensure that \textit{templates} provide sufficient documentation to users.

\subsubsection{Evaluating Guidelines/Pitfalls Adherence through LLMs}
After defining the guidelines and pitfalls, we explored whether LLMs could support practitioners in preliminarily assessing repository adherence to them.
For that, we used \texttt{GPT-5.2} with web search capabilities to evaluate a subset of projects previously analyzed in RQ3 (Section~\ref{sec:rq3}), providing each guideline/pitfall and its description as input.
A co-author not involved in defining or validating the guidelines manually evaluated the same projects, enabling us to compare human and LLM assessments.
The total cost of the LLM evaluation was approximately USD 1.40.
The comparison shows moderate agreement between human and LLM evaluations, with Cohen's $\kappa=0.46$, accuracy of 0.80, and macro F1-score of 0.73.
These results suggest that LLMs can reasonably approximate human judgments when assessing adherence to the proposed guidelines and pitfalls; however, human validation remains necessary for more interpretative criteria.
Agreement was higher for structural or easily observable criteria, such as G1 (software engineering practices) and P4/P5 (mixing technology and template concerns, and empty templates) ($\kappa=0.60$--$0.77$).
In contrast, lower agreement was observed for more interpretative criteria, G2 (documentation quality) and G7 (community engagement), which require deeper contextual understanding and subjective interpretation.
Overall, these results indicate that LLMs can serve as useful assistants for preliminary template evaluation, while human validation remains important for criteria requiring deeper contextual interpretation.

\section{Discussion}
\label{sec:discussion}

\ready{Our findings suggest that templates should be understood as upstream reusable artifacts rather than simple project starters.
As reported in RQ2 (Section~\ref{sec:rq2_results}), templates may exhibit multiple quality issues, including code smells, bugs, and vulnerabilities.
Since repositories instantiated from templates are not automatically synchronized with their original \textit{template} repositories by default, issues present at instantiation time may persist downstream, while later fixes are not automatically propagated.
If left unresolved, these inherited issues can contribute to template-specific technical debt over time~\cite{molnar2020long,digkas2020can}.}

\subsubsection*{Template Consumers and Authors}
\ready{For template consumers, our results indicate that adoption decisions should go beyond signals such as stars, forks, or ownership.
Although these signals may suggest popularity or organizational support, RQ2 shows that their relationship with quality issues is ecosystem-dependent.
Consumers should therefore inspect dependency freshness, documentation quality, automation support, supported versions, and known maintainability/security issues before adoption.
For template authors, our findings suggest that templates should be maintained as reusable products rather than static examples.
Authors should provide clear versioning, deprecation/status signals, migration guidance, and separation between reusable scaffolding and full implementations.
Finally, since templates can be copied across projects and used as learning resources, poor template design may normalize inadequate practices, increase adaptation effort, and amplify local maintenance problems across broader ecosystems.}

\ready{Although our primary classification focuses on served domains rather than an explicit taxonomy of template purposes, our qualitative analysis suggests that templates play multiple roles in practice.
Some templates act as educational or tutorial scaffolds, helping newcomers learn technologies and adopt good development practices.
Others serve as framework starters or organizational boilerplates, aiming to standardize project setup, dependencies, automation, and release practices.
Such a variation in purpose helps to explain why documentation, versioning, and separation between template and implementation concerns are central to our guidelines and to the pitfalls.}

\subsubsection*{Researchers} When selecting and evaluating repositories from GitHub, researchers should be aware of the potential inclusion of templates in their samples.
Suppose the goal is to analyze the characteristics or practices of actual software development, then templates should be filtered out, as they do not represent complete or functional projects but rather scaffolds intended for future development.
Failing to identify and exclude them may introduce bias in empirical analyses, especially those examining development activity, team collaboration, or software evolution.

\subsubsection*{GitHub Maintainers} To facilitate future studies on templates, it is essential that GitHub provides better support for identifying and collecting templates.
Currently, the search API allows users to query exclusively for templates, although we could not find official documentation. 
Without such support, researchers could adopt keyword heuristics, as we initially did here. 
However, such a decision can negatively affect dataset quality and introduce bias in empirical studies.
Providing dedicated filters or metadata for templates would improve the reproducibility and accuracy of research that depends on large-scale repository mining.

\section{Related Work}
\label{sec:related-work}

\subsection{Template Repositories}
\label{subsec:template-repos}

Some studies primarily focused on using and evaluating the adoption of templates as a tool to promote and support students in courses \cite{van2024introduction, beckman2021implementing, snowberger2023workflow}. 
For example, Smith et al.\cite{smith2024software} explored how GitHub templates can facilitate software engineering education by providing a repository template as a structured starting point for student projects.
For that, the authors propose a course model based on GitHub templates to scaffold learning activities, streamline project setup, and standardize evaluation.
They report that the adoption of templates effectively reduced configuration overhead, encouraged good development practices, and supported reproducible coursework outcomes.
Such an educational perspective complements our results presented here, illustrating one of the supported domains behind template creation (\textit{Education}).

Similarly, Tu et al. \cite{tu2022github} investigate the role of GitHub in supporting structured project development in the educational context.
Specifically, the authors create repositories using them as templates to support students in initializing their projects with standardized structures, predefined configurations, and consistent development practices.
Although these repositories were not formally labeled as \textit{templates}, their purpose aligns with the fundamental idea of GitHub templates, providing reusable foundations that ensure consistency and reduce configuration effort. 
Such a pedagogical perspective reinforces one of the main motivations behind template creation observed in our study: lowering entry barriers and promoting uniform project organization.

While the previous studies focus on the educational domain, our findings demonstrate that similar principles of reusability and standardization also drive template adoption across industrial and open-source domains.
Together, these studies suggest that GitHub templates are evolving into important vehicles for both learning and professional software engineering practice, bridging the gap between scaffolding for education and reuse in real-world development.

\ready{Beyond template-based studies, previous work covering GitHub README files provides an important and relevant perspective on understanding reusable scaffolding artifacts. 
In this context, README files have been studied as a central artifact for communicating project goals, installation steps, usage examples, contribution expectations, and quality-related information~\cite{prana2019categorizing,liu2022readme,wang2023study,gaughan2025introduction}. 
These studies show that repository documentation is closely related to how projects are understood, adopted, and maintained. 
Such a perspective is particularly relevant for template repositories, since templates must not only provide code but also explain how the repository should be instantiated, customized, and evolved. 
While these previous studies analyze documentation practices in general GitHub repositories, our work complements them by studying repository-level templates as reusable artifacts and deriving guidelines that include documentation, maintenance, and reuse-specific concerns.
}

\subsection{Collaborative Development and Structured Templates}
\label{subsec:collaborative-dev}

Over time, GitHub has become the largest ecosystem for collaborative software development, being extensively studied across multiple dimensions.
For example, previous studies have investigated pull-based development workflows \cite{gousios2014exploratory}, fork-based reuse \cite{jiang2017and}, and project popularity \cite{borges2016understanding}. 
Previous studies have also investigated software evolution in depth, covering how collaborations occur and diverge with each other \cite{da2024detecting, da2022build}, quality and maintenance characteristics evolve \cite{lenarduzzi2019technical}, and the impact of adopting automation tools \cite{kinsman2021software}. 
Researchers have also proposed guidelines for conducting empirical studies using GitHub data \cite{kalliamvakou2014promises, kalliamvakou2016depth}, while other studies have examined best practices and pitfalls related to communication and collaboration within GitHub projects \cite{hata2022github}.
However, most of this literature focuses on traditional repositories, exploring different mechanisms, such as forks and pull-requests (PRs), leaving templates largely unexplored.

With respect to repository management, GitHub provides multiple features exploring the general concept of \textit{templates}, including support for issue reports, pull requests, and GitHub Action workflows \cite{li2022follow}. 
Prior research has examined the adoption and impact of such templates in contribution-level contexts. 
For example, S\"ul\"un et al. \cite{sulun2024empirical} analyzed 350 templates from 100 projects while collecting insights from practitioners, showing that structured issue templates reduce reporting time and improve information quality. 
Similarly, Zhang et al. \cite{zhang2022consistent} conducted an empirical study on pull request templates, observing that structured templates enhance review consistency and reduce the time and effort required during code review. 
However, these studies focus on contribution-level templates rather than repository-level templates, as we investigate here.

\section{Threats to Validity}
\label{sec:threats}

\textit{Construct to Validity.}
As previously informed, the GitHub API does not officially support searching for templates.
To mitigate this limitation, we initially employed a keyword-based heuristic as a proxy for identifying potential templates.
However, during further experimentation, we discovered that including the parameter \texttt{template:true} in the query reliably returned repositories explicitly marked as templates, which we adopted to strengthen the accuracy and completeness of our dataset.
Nevertheless, to ensure consistency and reduce potential bias, we restricted our analysis to repositories explicitly labeled as templates in our final dataset.

Our taxonomy of domains was defined through manual analysis conducted by a single researcher, possibly introducing classification bias.
Aiming to mitigate this risk, two independent researchers performed additional analyses, including validation against LLM-generated categorization, to improve consistency and robustness.
When comparing human- and LLM-based domain classifications for RQ1, we observed some disagreements. By manually inspecting them, most conflicts occurred in repositories crossing multiple domains, where classification boundaries are inherently ambiguous rather than completely incorrect. 
Although minor differences in categorization were observed, the overall domain distribution and reported trends remained consistent. 
Therefore, we believe that these discrepancies may not directly affect the validity of the reported results.

\textit{Internal Validity.} When investigating the factors associated with quality issues in RQ2, we highlight that the associations should not be interpreted causally. 
Repository-level characteristics such as recent activity, size, and popularity may be interrelated, and confounding factors, like the domain complexity or framework usage, may influence the observed relationships. 
Although we controlled for size and assessed multicollinearity, unobserved factors may still affect the results.

\textit{Threats to Reliability.} Although qualitative analysis inherently involves interpretation, we mitigated individual researcher bias through independent analysis and consensus-based consolidation, reducing the influence of personal perspectives on the reported insights.
\ready{Our LLM-based domain classification also used a single run per repository. 
While we used a low temperature setting and manually validated a representative sample, we acknowledge that we do not directly measure run-to-run variability in the generated labels.} 

\textit{Conclusion Validity.} To assess potential multicollinearity among predictors, we computed variance inflation factors (VIF) for all models. 
All variables exhibit VIF values below commonly accepted thresholds (around 1), indicating that multicollinearity does not materially affect our estimates and that observed changes in coefficient significance reflect conditional effects rather than unstable regression artifacts.

\textit{External Validity.} By explicitly relying on repositories marked as \textit{templates}, we may have excluded repositories that serve similar purposes but are not properly masked as such.
This limitation is relevant in our context, considering that the template feature was only introduced by GitHub in 2019 \cite{githubTemplates}.
Although our study covers the five most used programming languages on GitHub, our results may not generalize to templates written in other programming languages.

\section{Conclusion}
\label{sec:conclusion}

Templates represent a relatively recent mechanism for sharing code on GitHub, focusing on providing an initial structure that practitioners can learn from and extend when building new solutions.
In this study, we aimed to further understand the domains that templates support, the factors associated with maintenance and quality issues, and the guidelines and pitfalls practitioners should consider when proposing or selecting templates. 
To achieve this goal, we conducted a large-scale empirical study of GitHub templates across the five most popular programming languages on the platform (\TypeScript,\Python, \JavaScript, \Java, and \CSharp).

Our results show that templates span diverse domains, with \textit{Web Development} being the most prevalent, a pattern consistently observed across programming languages. Regarding maintenance and quality issues, we observe substantial variability, with repositories exhibiting different levels and densities of issues. 
These findings indicate that maintenance and quality issues in templates are strongly ecosystem-dependent rather than uniformly distributed. 
Finally, we provide a set of actionable guidelines and recurring pitfalls, highlighting how design, documentation, automation, versioning, and governance decisions in templates can directly influence their reusability and the quality of projects instantiated from them.
 
By providing the first large-scale empirical analysis of GitHub \textit{templates}, this study contributes to a deeper understanding of how reusable scaffolding repositories shape modern software development practices. 
For researchers, our findings open avenues for future work on the long-term impact of templates on software quality, the evolution of template families, and the role of templates in disseminating best practices across ecosystems. 
Furthermore, we also report implications for the adoption of templates in empirical research and discuss the need for improved support from GitHub maintainers to facilitate mining template-related information.

\section*{AI-Assisted Content Disclosure}

LLMs were used to assist the authors with grammar revision and support for some scripts used in this study. All outputs were reviewed and validated by the authors. 

\section*{Acknowledgments}
This work was partially supported by the Natural Sciences and Engineering Research Council of Canada, the Canadian Institute for Advanced Research, and the Canada Research Chairs Program.

\bibliographystyle{IEEEtran}
\bibliography{sample-base}
\end{document}